\documentclass[12pt,a4paper]{article}
\usepackage[a4paper,margin=1in]{geometry}
\usepackage{times}
\usepackage{setspace}
\usepackage{hyperref}
\usepackage{titlesec}
\usepackage{graphicx}
\usepackage{amsmath, amssymb}
\usepackage{booktabs} 

\titleformat{\section}{\large\bfseries}{\thesection.}{0.5em}{}
\titleformat{\subsection}{\normalsize\bfseries}{\thesubsection.}{0.5em}{}

\title{Einstein’s 1935 Letters to Schrödinger and Popper and the Boundaries of the PBR $\psi$-Epistemic Framework}

\medskip

\author{Galina Weinstein\thanks{The Department of Philosophy, University of Haifa.}}
\date{}

\begin{document}
\maketitle
\doublespacing

\begin{abstract}
Einstein’s 1935 critique of quantum mechanics is often associated with the Einstein--Podolsky--Rosen (EPR) argument, yet his private correspondence from that year reveals a more exact conceptual structure guiding his claim that the $\psi$-function is incomplete. This paper reconstructs Einstein’s reasoning in his letters to Schrödinger and Popper and examines how it engages, and fails to engage, with contemporary $\psi$-ontic/$\psi$-epistemic distinctions. Recent scholarship, most notably by Ben{-}Menahem, has interpreted Einstein as an early representative of the modern $\psi$-epistemic tradition within the Harrigan--Spekkens ontological-models framework and the Pusey--Barrett--Rudolph (PBR) theorem. I argue, however, that this retrospective classification is undermined by Ben{-}Menahem’s own distinction between realist and radical epistemic interpretations: Einstein’s 1935 view lacks the structural assumptions---a defined ontic state space, preparation distributions, and overlap structure---required for membership in the HS/PBR class of $\psi$-epistemic models. Any such identification, therefore, requires importing formal machinery foreign to Einstein’s original argument.
\end{abstract}

\section{Introduction}

Einstein’s 1935 critique of quantum mechanics is often identified with the published Einstein--Podolsky--Rosen (EPR) paper \cite{EPR}. Yet the conceptual core of that critique is frequently clearer in Einstein’s private correspondence than in the canonical article itself. In particular, two letters from 1935—one to Erwin Schrödinger (19 June), written shortly after the appearance of EPR, and one to Karl Popper (11 November)—frame Einstein’s concerns about completeness in two distinct registers, and together illuminate the structure of his argument more sharply than the published paper alone.

The present study reconstructs Einstein’s reasoning in these letters and situates that reconstruction within the broader interpretive landscape shaped by contemporary distinctions between $\psi$-ontic and $\psi$-epistemic models. This landscape has been strongly influenced by the Harrigan--Spekkens (HS) ontological-models framework \cite{H-S} and the Pusey--Barrett--Rudolph (PBR) theorem \cite{PBR}, which articulate formal criteria for when a quantum state is to be regarded as ontic or epistemic. In recent years, these developments have prompted attempts to locate historical figures—Einstein among them—within the resulting taxonomy.

One prominent example is the work of Yemima Ben{-}Menahem \cite{YBM-3, YBM-4, YBM, YBM-1, YBM-2}, who draws on the 1935 correspondence, particularly the Popper letter, to place Einstein within a broadly “epistemic’’ or ensemble-based tradition. Because this interpretation relies in part on the conceptual resources of the HS/PBR framework, it raises several methodological questions about how twentieth-century debates should be related to contemporary classificatory schemes. A closer examination of Einstein’s original texts offers an opportunity to reassess this relationship.

The paper proceeds as follows. Section \ref{1} reconstructs the structure of Einstein’s reasoning in his 1935 letters to Schrödinger and Popper, identifying the conceptual elements that guide his analysis of completeness. Section \ref{2} provides an outline of the HS ontological-models framework and the PBR theorem, with particular attention to how PBR themselves compare their result with Einstein’s argument. Section \ref{3} turns to Ben{-}Menahem’s application of the PBR taxonomy to Einstein and considers how Einstein’s correspondence fits, or fails to fit, the assumptions implicit in that classificatory approach.

\section{Einstein's Letters to Schrödinger and Popper} \label{1}

\subsection{Einstein’s 19 June 1935 Letter to Schrödinger} 

Einstein’s letter of 19 June 1935 to Schrödinger—written only weeks after the publication of the Einstein–Podolsky–Rosen (EPR) paper \cite{EPR}—is among his most direct and revealing statements on the incompleteness of quantum mechanics. Although the published EPR article \cite{EPR} has often been read as advancing a criticism of the wavefunction’s completeness, the private letter makes clear that Einstein understood the decisive issue to be \emph{ontological}: the coordination between real physical states and their theoretical representation. That coordination, he insists, is constrained by a separability requirement. In what follows, I reconstruct his reasoning directly from the German text. While Arthur Fine \cite{Fine} and Don Howard \cite{How} have examined this material, my aim here is to offer a distinct interpretation.

Einstein begins by distancing himself from the published EPR argument. Podolsky drafted the paper, he notes, and in the process, “the main point is, so to speak, buried under scholarship.”%
\footnote{Translation from the German is my own \cite{Fine, How, Me}.}
The letter, then, is Einstein’s effort to restate the core of the argument clearly.

He next identifies the underlying philosophical difficulty: physics aims to describe reality, yet we access reality only through our theories. “Physics describes ‘reality,’ but we do not know what ‘reality’ is; we know it only through the physical description!”%
\footnote{Translation from the German is my own \cite{Fine, How, Me}.}
For Einstein, this is not an invitation to instrumentalism. Instead, it underscores that a successful formalism may still fail to represent real states adequately.

Einstein then draws a basic distinction: “All physics is a description of reality; but this description can be ‘complete’ or ‘incomplete’.”%
\footnote{Translation from the German is my own \cite{Fine, How, Me}.}
To illustrate what “incomplete” means, he introduces a classical toy model. A ball is in one of two boxes; before opening a box, we assign probability $\dfrac{1}{2}$ to each. This, he stresses, is \emph{not} a complete description: “A complete statement is: the ball is in the first box (or it is not).”\footnote{Translation from the German is my own \cite{Fine, How, Me}.} Here probability appears only to clarify completeness: statistical statements reflect ignorance of determinate facts arising from uncontrolled external factors in observation. The example sets a benchmark: a complete description must reflect what is the case, not merely what is expected. This classical illustration is not meant as an interpretation of quantum probabilities; it serves only to fix what Einstein means by a “complete” description.

Turning to quantum mechanics, Einstein writes that the theory “uniquely associates” a wavefunction with the real state of a system. 
Crucially, he introduces this as the \emph{completeness claim to be tested}. If a one-to-one correspondence between $\psi$ and the real state cannot be maintained, the description is incomplete: “If such an interpretation cannot be carried out, I call the theoretical description ‘incomplete’.”%
\footnote{Translation from the German is my own \cite{Fine, How, Me}.} 

Quantum mechanics, he argues, fails this test. The theory assigns multiple legitimate wavefunctions to the same real state of subsystem $B$, violating the bijective coordination required for completeness.

Einstein then presents a sharp formulation of the EPR scenario. Two systems $A$ and $B$ interact, separate, and are described by a joint wavefunction $\psi_{AB}$. 
Different measurements on $A$ produce different conditional wavefunctions for $B$, all derived from the same $\psi_{AB}$, even though (by hypothesis) the real physical state of $B$ remains fixed after separation.

At this point, Einstein invokes the decisive additional assumption. With characteristic humor, he remarks that one cannot “overcome the Talmudist"—the one demanding that every inference be spelled out—without an extra principle. As Howard notes \cite{How}, later correspondence shows that Einstein had Niels Bohr in mind. To block the Talmudist objection, Einstein states the \emph{Trennungsprinzip} (separation principle): the state of $B$ is independent of what is done to the spatially separated $A$ \cite{Fine, How, Me}.

Once separability is imposed, the formalism yields the core contradiction:  
for the same real state of $B$, quantum mechanics assigns multiple equally legitimate $\psi_B$. 
Thus, $\psi$ cannot be a complete description. If it were, different choices of measurement on $A$ would have to change the real state of $B$ instantaneously—an option Einstein rejects: “The real state of $B$ cannot depend on what kind of measurement I perform on $A$.”%
\footnote{Translation from the German is my own \cite{Fine, How, Me}.}
Einstein concludes that the non-uniqueness of $\psi_B$ is incompatible with any one-to-one representational claim.
He urges that quantum mechanics be regarded as incomplete until a more adequate theory is found.

\subsection{Einstein’s Technical Derivation in the 19 June 1935 Letter}

Unlike his more informal correspondence with Popper, Einstein’s long letter to Schrödinger of 19 June 1935 contains a compact but fully explicit mathematical derivation \cite{Me}. Einstein uses the linear structure of Hilbert space to make an ontological point: the wavefunction cannot be in one-to-one correspondence with the real physical state.

Einstein considers two systems, $A$ and $B$, which interact, separate, and are thereafter described by a joint wavefunction $\Psi_{\text{AB}}(x_1,x_2)$. He begins by expanding the total state in the eigenfunctions of an observable $\alpha$ associated with subsystem $A$. He writes \cite{Me}:
\begin{equation} \label{Psi}
\Psi_{\text{AB}}(x_1,x_2)
  = \sum_{n} c_{mn}\,\Psi_m(x_1)\,\chi_n(x_2). 
\end{equation}
As Einstein presents it, the index $m$ is held fixed; equation \eqref{Psi} represents the $m$-th row of coefficients in the full bipartite expansion. The coefficients $c_{mn}$ may therefore be viewed as the entries of that row of a matrix, with $n$ labeling the eigenfunctions $\chi_n$ of subsystem $B$. This is the linear-algebraic decomposition of a generally entangled bipartite state, and it already shows how the joint description depends on the choice of basis for $A$.

Einstein now considers performing an $\alpha$-measurement on subsystem $A$. If the eigenvalue corresponding to $\Psi_m$ is obtained, he writes that the description of subsystem $B$ reduces to \cite{Me}:
\begin{equation} \label{eq:einstein-psiB}
\Psi_{\text{B}}(x_2)
  = \sum_{n} c_{mn}\,\chi_n(x_2). 
\end{equation}
This expression is precisely the $m$-th row of amplitudes $(c_{m1},c_{m2},\ldots)$ from the expansion in \eqref{Psi}. It is the conditional wavefunction assigned to $B$ given the outcome $m$ has been obtained on $A$. Nothing on the side of $B$ has changed physically; all that has happened is that $A$ has been measured after separation.%
\footnote{This \emph{is ontological}, a metaphysical claim \emph{about the world, not about knowledge}. An epistemic version of this claim would be something like: Our knowledge or information about $B$ has changed after measuring $A$, even if the physical situation of $B$ has not changed.}

Einstein emphasizes that merely by choosing which observable to measure on the already separated system $A$, the theory assigns a \emph{different} wavefunction to the same physical object $B$. This is the core of the problem.

To sharpen the point, Einstein repeats the construction for a \emph{different} observable $\beta$ of $A$. In this case, he expands the same total state in a different eigenbasis $\{\Psi_m'(x_1)\}$, and here he explicitly writes the full double sum \cite{Me}:
\begin{equation} \label{eq:einstein-beta-expansion}
\Psi_{\text{AB}}(x_1,x_2)
  = \sum_{n} c'_{mn}\,\Psi'_m(x_1)\,\chi_n(x_2). 
\end{equation}
From this second expansion, the corresponding conditional wavefunction of $B$ after obtaining the $m$-th $\beta$-outcome is \cite{Me}:
\begin{equation} \label{2a}
\Psi'_{\text{B}}(x_2)
  = \sum_{n} c'_{mn}\,\chi_n(x_2). 
\end{equation}
Thus, measuring $\alpha$ on $A$ yields $\Psi_\text{B}$, while measuring $\beta$ on $A$ yields $\Psi'_\text{B}$. But $A$ and $B$ are spatially separated and non-interacting. Nothing physical happens to $B$; only the choice of basis (and hence the measurement) on the remote system $A$ changes the wavefunction assigned to $B$.

Einstein concludes that $\Psi_\text{B}$ and $\Psi'_\text{B}$ must correspond to the same real physical state of $B$ after separation. Yet, the theory yields different wavefunctions for that state depending on the measurement performed on $A$. Therefore, the wavefunction cannot be a complete description of the real state (see German text in \cite{Me}).

This is the technical core of Einstein’s incompleteness argument. If $\Psi$ were complete, then the real state of $B$ would have to depend on which measurement is performed on distant $A$, a consequence Einstein rejects by appeal to the \emph{Trennungsprinzip}. The non-uniqueness of $\Psi_B$ therefore shows that the wavefunction cannot be in one-to-one correspondence with the real physical state. 

Einstein’s derivation already contains the key structural features that Schrödinger would later single out in his 1935 trilogy on \emph{Verschränkung} \cite{cat}.

First, the joint state $\Psi_{\text{AB}}(x_1,x_2)$ in equations \eqref{Psi} and \eqref{eq:einstein-beta-expansion} is generically non-factorizable: the coefficients $c_{mn}$ and $c'_{mn}$ cannot, in general, be written as products $a_m b_n$ or $a'_m b_n$. Thus:

\begin{equation}
\Psi_{\text{AB}} \neq \Psi_\text{A} \otimes \Psi_\text{B},    
\end{equation}
which is precisely the condition that the pure bipartite state be entangled in Schrödinger’s sense.

Second, Einstein's equations \eqref{eq:einstein-psiB} and \eqref{2a} exhibit the \emph{non-uniqueness} of the conditional state of the subsystem. From a single fixed joint state $\Psi_{\text{AB}}$ one obtains, by choosing different observables $\alpha$ and $\beta$ on $A$, two different conditional wavefunctions, which, according to Einstein, nevertheless refer to the same real physical state of $B$ after separation. This is the phenomenon that Schrödinger later characterizes as the characteristic trait of entangled systems: best possible knowledge of the whole does not fix a unique pure-state description of the parts.

Third, Einstein’s construction already contains what Schrödinger would later describe as the \emph{steering} of the distant system: by selecting which observable is measured on $A$, one can bring about, within the theory, different pure-state assignments $\Psi_\text{B}$ or $\Psi'_\text{B}$ for $B$, without any physical intervention on $B$ itself. The choice of measurement on $A$ thus "steers" the theoretical state of $B$ through a family of alternative wavefunctions.

Finally, Einstein states that the subsystem's conditional states are \emph{basis-dependent}. The same $\Psi_{\text{AB}}$ admits many inequivalent decompositions into product terms, such as \eqref{Psi} and \eqref{eq:einstein-beta-expansion}, each tied to a different choice of observable on $A$. Schrödinger’s later discussion of entanglement as essentially tied to such basis-dependent decompositions of the joint state is already anticipated here in Einstein’s letter.

Taken together, these points show that the 19 June 1935 letter not only contains Einstein’s incompleteness argument, but also anticipates in technical form several of the central features of entanglement that Schrödinger would later emphasize in his cat paper \cite{cat} and the related 1935-1936 articles.

\subsection{Einstein’s Letter to Popper (11 November 1935)} 

In contrast to Einstein’s contemporaneous letter to Schrödinger (19 June 1935), which presents a technical and explicitly ontological analysis of the wavefunction in the EPR setting, the Popper letter adopts a more philosophical and methodological register. The difference reflects its intended audience: Popper was a philosopher of science, and Einstein therefore formulates the issues in terms of interpretation, the meaning of probability in physics, and the relation between determinism and statistical laws.

In the letter to Popper, Einstein’s discussion remains deliberately intuitive, and even when he invokes contemporary figures, the references are conceptual rather than technical. He mentions Heisenberg by name—"Heisenberg flirts with such a conception, without adopting it consistently"%
\footnote{Translation from the German is my own \cite{Popper}.}
—but only in the course of a reflection on the statistical character of quantum mechanics. Crucially, Einstein does not invoke the "Unschärferelation," makes no mention of “Unsicherheitsrelationen,” and nowhere appeals to the commutation structure that underlies Heisenberg’s uncertainty relations. The Popper letter contains no trace of the operator-based uncertainties that form the mathematical backbone of the EPR paper. Its treatment of the mutual exclusivity of precise position and momentum determinations is qualitative, cast entirely in terms of wave packets, knowledge, and disturbance, rather than in terms of noncommuting operators.

This absence is significant. In the EPR paper, the impossibility of simultaneous sharp values is described mathematically in the non-commutativity of $P$ and $Q$, and the argument proceeds by working out the implications of that formal structure for the notion of completeness. In the Popper letter, the same physical conclusion is expressed \emph{without the Heisenberg machinery and without any algebraic reference to the canonical commutation relations}. Einstein effectively translates the EPR argument into the language of physical intuition, rather than repeating the formal reasoning he, Boris Podolsky, and Nathan Rosen had already published.
Yet Einstein does something in the Popper letter that is not present in the EPR paper in this explicit form: he appeals directly to what, in his letter to Schrödinger written a few months earlier, he called the \emph{Trennungsprinzip}, the separation principle.

Einstein opens diplomatically, acknowledging substantial agreement with Popper’s general aims: "I have examined your paper and, for the most part, I am in agreement."%
\footnote{Translation from the German is my own \cite{Popper}.}
This signals that the reply will be corrective rather than adversarial. Einstein treats Popper as a serious interlocutor but aims to clarify where Popper’s reasoning misfires in relation to quantum mechanics.

In his letter to Popper, Einstein first separates two issues: 

1) Limited predictive precision at the atomic scale is unsurprising: “I consider it trivial that one cannot make arbitrarily precise predictions in the atomic domain \dots”%
\footnote{Translation from the German is my own \cite{Popper}.} For Einstein, the mere existence of statistical predictions does not decide whether quantum probabilities reflect ignorance of determinate microfacts or genuine indeterminism. 

2) He then identifies the central interpretive fork: "One cannot say whether the statistical character of our experimental findings is brought about, according to today’s quantum theory, \dots while the systems themselves \dots behave deterministically in themselves."%
\footnote{Translation from the German is my own \cite{Popper}.} 
Here, Einstein makes explicit that the issue is ontological: quantum statistics may express ignorance about an underlying deterministic microdynamics. Positivist appeals to measurement limits do not settle the matter.

Einstein then presents Popper with a compact, non-technical version of the EPR structure. Consider a bipartite system $A+B$ that interacts and then separates. The wavefunction before the interaction is known, and the Schrödinger equation yields the post-interaction wavefunction: "The $\psi$-function of the total system before the interaction is known. The Schrödinger equation then yields the $\psi$-function of the total system after the interaction."%
\footnote{Translation from the German is my own \cite{Popper}.}

Once a complete measurement is performed on $A$, quantum mechanics assigns a conditional state to $B$—and, crucially, different choices of complete measurement on $A$ lead to different conditional $\psi_B$: "Now, let a complete measurement be performed on subsystem $A$ \dots. Quantum mechanics then provides the $\psi$-function for subsystem $B$, and indeed a different one depending on the choice of measurement made on $A$."%
\footnote{Translation from the German is my own \cite{Popper}.}

Einstein then inserts the decisive premise: separability/locality. The real physical state of $B$ cannot depend on what kind of measurement is freely chosen on $A$: "But since it is unreasonable to assume that the physical state of $B$ depends on what kind of measurement I perform on the system $A$, which is separate from it, \dots"%
\footnote{Translation from the German is my own \cite{Popper}.}

From this, Einstein derives the central conclusion:
"\dots this means that two different $\psi$-functions correspond to the same physical state of $B$. Since a complete description of a physical state must necessarily be a unique description, the $\psi$-function cannot be regarded as a complete description of the state."%
\footnote{Translation from the German is my own \cite{Popper}.}

Here, the incompleteness verdict is articulated as a methodological constraint: locality and quantum predictions together force non-uniqueness of $\psi$ for a single real state. The Popper letter, therefore, presents a philosopher’s formulation of precisely the same structure that the Schrödinger letter articulates technically.

Einstein then bolsters the argument with a counterfactual-definiteness claim. Suppose a quantity can be predicted with certainty by choosing an appropriate measurement on $A$. In that case, $B$ possesses the corresponding property: "One can hardly avoid the view that system $B$ actually has a definite momentum and a definite coordinate. For whatever I can predict at will must also exist in reality."%
\footnote{Translation from the German is my own \cite{Popper}.}

Popper had argued that deterministic theories cannot yield statistical laws. Einstein explicitly rejects this: "I want to say once again that I do not consider your claim that no statistical statements can be derived from a deterministic theory to be correct."%
\footnote{Translation from the German is my own \cite{Popper}.}

He illustrates the point with classical examples, gas theory and Brownian motion, where deterministic microdynamics lead to statistical laws solely because initial conditions are unknown: "What is essential is only that I do not know the initial state, or do not know it precisely!"%
\footnote{Translation from the German is my own \cite{Popper}.}

The philosophical moral is clear: quantum statistics do not entail indeterminism. They may reflect incomplete knowledge of underlying microfacts not represented in $\psi$. The burden of proof lies with those who deny this possibility.

The lesson Einstein draws is one of \emph{ontological incompleteness}. Quantum statistics do not demonstrate that physical processes lack determinate causes; they show only that the quantum state is not a full representation of the underlying microconditions. As in classical kinetic theory, probabilities may arise from incomplete access to deeper, determinate physical facts. Thus, the incompleteness lies in the ontology represented by the theory, not in our subjective knowledge.

Taken together, the core of the Popper letter is this: probability does not decide between determinism and indeterminism; locality implies that $\psi$ cannot be complete; and deterministic theories can yield statistical laws without contradiction.

\subsection{Schrödinger: the technical EPR; Popper: the philosophical EPR} \label{pos}

The difference between Einstein’s two 1935 letters lies not in their conclusions but in their presentation, tone, and intended audience.

The Schrödinger letter develops a technical, ontological contradiction internal to quantum formalism. It presupposes familiarity with entangled $\psi$–functions, invokes the separation principle as a precise physical postulate, and derives incompleteness from the non-uniqueness of $\psi_B$ under locality.

The Popper letter, by contrast, articulates a philosophical clarification of probability, determinism, and the logic of explanation. Einstein’s critical remark about the “modische, positivistische Kleben am Beobachtbaren’’ (“fashionable, positivistic clinging to the observable’’) signals a broader methodological critique rather than a technical proof.

In the Schrödinger letter, the central thesis is ontological non-uniqueness: the same real state of $B$ corresponds to multiple $\psi_B$.  
In the Popper letter, the target of the argument is Popper’s methodological inference from statistical formalism to indeterminism. Einstein’s response highlights classical cases in which deterministic systems yield statistical predictions.

Thus, the Popper letter represents no change in Einstein’s position. It restates, in philosophical idiom, the same locality-based non-uniqueness argument found in the Schrödinger correspondence. Both letters express a single, continuous position: \emph{the argument is ontological, not epistemic}, and it aims to show that under locality, the wavefunction cannot be a complete description of an individual physical system.

\section{The PBR Ontological Framework} \label{2}

\subsection{The Harrigan--Spekkens (HS) \texorpdfstring{$\psi$-Epistemic View}{psi Epistemic View}}

The Pusey--Barrett--Rudolph (PBR) theorem is formulated within a very general class of ontological models that attempt to explain quantum probabilities by positing that each individual system possesses an underlying \emph{real} physical state. 
The PBR theorem is formulated within the general ontological-models framework developed by Nicholas Harrigan and Robert Spekkens (HS), and it targets in particular the class of so-called $\psi$-epistemic models in their sense \cite{H-S}. This picture is explicitly inspired by the way classical statistical mechanics treats macrostates.

In classical statistical mechanics, a macrostate (temperature, pressure, etc.) does not specify the precise microstate; \emph{many} different microstates are compatible with the \emph{same} macrostate; and the macrostate represents \emph{incomplete} information about the true underlying situation.

HS apply this analogy to quantum theory. A quantum state $\psi$ plays the role of a classical macrostate; the underlying \emph{ontic state} $\lambda$ plays the role of the full microstate; and preparing $\psi$ corresponds to selecting a probability distribution over the possible $\lambda$'s. Thus, on this view, $\psi$ represents information rather than physical reality itself. This is the conceptual framework that HS formalize, and which PBR investigate.

PBR's central contribution is to show that this analogy fails under extremely mild assumptions \cite{PBR}. In classical statistical mechanics, the same microstate may lie inside different macrostates, and such \emph{overlap} is expected because macrostates are coarse-grained descriptions. If quantum states behaved analogously, then two distinct wavefunctions could overlap in their distributions over ontic states. That is, the \emph{same} real physical situation could be compatible with \emph{different} wavefunctions. In \emph{a $\psi$-epistemic model} one might have: 

\begin{equation} \label{Psilambda}   
\psi_0 \longrightarrow \text{sometimes produces }\lambda^*, \qquad
\psi_1 \longrightarrow \text{sometimes also produces }\lambda^*.
\end{equation}
This possibility of overlap is precisely what allows $\psi$ to be purely informational: the same underlying reality is compatible with multiple $\psi$'s.

In a $\psi$-epistemic interpretation, the wave function is not regarded as a physical entity. What is ordinarily called collapse must therefore be interpreted as a change in the observer’s knowledge, not as a change in the system itself. In this sense, wave-function collapse is directly analogous to Bayesian updating in classical probability theory.

In Bayesian inference, when new information $D$ becomes available, a rational agent updates a prior probability distribution $P(H)$ over hypotheses $H$ by applying Bayes’s rule:
\begin{equation}
P(H \mid D) = \frac{P(D \mid H)\,P(H)}{P(D)}.
\end{equation}
A $\psi$-epistemic interpretation treats quantum measurement in exactly this way. A measurement outcome provides new information about $\lambda$, and the agent updates the probability distribution accordingly.

Before measurement, the wave function typically assigns nonzero weights to multiple possibilities. For example, a state:
\begin{equation}
|\psi\rangle = \alpha |0\rangle + \beta |1\rangle
\end{equation}
is interpreted not as a literal superposition of physical properties, but as a probability distribution over states $\lambda$ compatible with outcomes $0$ or $1$. If the measurement yields the outcome $1$, the agent rules out all states $\lambda$ compatible only with outcome $0$ and assigns full weight to those compatible with outcome $1$. The post-measurement wave function is therefore nothing more than the \emph{posterior} distribution conditioned on the observed outcome. Symbolically:
\begin{equation}
\psi_{\text{after}} = \psi_{\text{before}} \quad \text{updated on the measurement result}.
\end{equation}
This is Bayesian conditionalization in quantum language.

A familiar classical analogy clarifies the point. When we draw a card from a deck, the probability distribution over the possible cards changes abruptly once we look at it, but the card itself has not changed; only our knowledge has. The $\psi$-epistemic picture claims that quantum collapse is of precisely this kind. The system is assumed to occupy some definite physical state $\lambda$ all along; the wave function describes incomplete knowledge; and measurement induces an update of that knowledge upon receiving new information. Collapse is therefore not a physical process but the agent’s rational updating of their probability assignments in light of the measurement result.

\subsection{The Pusey--Barrett--Rudolph (PBR) Theorem}
\label{PBRT}

PBR show that the relations \eqref{Psilambda} cannot be maintained. Their argument uses the following assumptions:
\begin{enumerate}
    \item \emph{First premise: Ontic realism:} 
    
    \noindent Each preparation results in an objective underlying state $\lambda$.
    \item \emph{Second premise: Preparation independence:} 
    
    \noindent Independently prepared systems have independent ontic states; i.e., the joint distribution over ontic states factorizes.
    \item \emph{Third premise: Correct quantum predictions:} 
    
    \noindent The ontological model reproduces quantum theory's predictions.
    \end{enumerate}
\emph{What the PBR argument rules out, under these assumptions, is precisely the $\psi$-epistemic possibility that distinct quantum states have overlapping supports in ontic state space, i.e.\ that their associated distributions $\mu_{\psi}(\lambda)$ assign nonzero probability to some of the same underlying states.}

\medskip

\emph{The second premise}: To assume that systems are prepared independently means that we have two (or more) preparation devices operating separately, with no causal coordination relevant to the experimental run. Each device prepares a system in a chosen quantum state, and the ontic state produced by one device does not influence, and is not influenced by, the ontic state produced by the other. If system $A$ is prepared in one lab and system $B$ in another, then the real physical state of $A$ should not be secretly correlated with the real physical state of $B$. Formally, the joint distribution over ontic states factorizes into the product of the individual distributions \cite{PBR}. Rejecting this assumption would require highly contrived hidden correlations between allegedly independent devices.

From the assumption of preparation independence, if two systems are prepared independently, there is a nonzero probability that their underlying ontic states both lie in the overlap region between the distributions associated with two distinct quantum states. The PBR construction then exhibits a measurement on the composite system whose quantum predictions are incompatible with any ontological model in which such overlapping ontic states occur. Hence, the mere existence of an overlap leads directly to a contradiction with quantum mechanics \cite{PBR}.

In the PBR construction, the measurement operators are defined by four vectors $|\xi_{ij}\rangle$, one for each possible outcome $(i,j)$. These vectors are chosen so that the product state $|\psi_i\rangle \otimes |\psi_j\rangle$ is \emph{orthogonal} to the corresponding measurement vector:
\begin{equation} \label{Orth}
\langle \xi_{ij} \,|\, \psi_i \psi_j \rangle = 0.    
\end{equation}
$\langle \xi_{ij} \,|\, \psi_i \psi_j \rangle$ is the inner product between \emph{the measurement state} $|\xi_{ij}\rangle$ and \emph{the system state} $|\psi_i \psi_j\rangle$.
Equation \eqref{Orth} defines orthogonality: the inner product is zero, meaning the two states \emph{have no overlap} in Hilbert space. Equation \eqref{Orth} means that each product state $|\psi_i \psi_j\rangle$ is orthogonal to exactly one measurement vector $|\xi_{ij}\rangle$, and therefore the probability of obtaining outcome $(i,j)$ is exactly zero for that preparation.

This has an immediate consequence for the probability of obtaining outcome $(i,j)$:
\begin{equation} \label{0}
p(i,j) = \bigl| \langle \xi_{ij} \,|\, \psi_i \psi_j \rangle \bigr|^{2} = 0.    
\end{equation}
This equation corresponds to the probability formula: 
\begin{equation*}
p=| \langle \text{measurement state} | \text{system state} \rangle |^2.    
\end{equation*}
Equation \eqref{0} gives probability $= 0$ for orthogonal states. Because the inner product is zero (no overlap) \eqref{Orth}, the square is also zero. 

Because the four product states are all distinct, and because the measurement is constructed so that each one is orthogonal to a different measurement vector: 
$|\psi_0\psi_0\rangle \perp |\xi_{00}\rangle$, 
$|\psi_0\psi_1\rangle \perp |\xi_{01}\rangle$, 
$|\psi_1\psi_0\rangle \perp |\xi_{10}\rangle$, 
$|\psi_1\psi_1\rangle \perp |\xi_{11}\rangle$, we get four different forbidden outcomes.

In an ontological model, a physical system possesses a real ontic state $\lambda$, and the quantum state $|\psi\rangle$ 
describes only an experimenter's \emph{information} about $\lambda$.  
Accordingly, each quantum state is associated with a probability distribution $\mu_\psi(\lambda)$ over the space of physical states.  
A $\psi$-epistemic model allows for distinct quantum states to have overlapping supports:
\begin{equation}
    \mathrm{supp}(\mu_{\psi_0}) \cap 
    \mathrm{supp}(\mu_{\psi_1}) \;\neq\; \varnothing .
\end{equation}
This means that the same underlying ontic state $\lambda$ may be compatible with either preparation $|\psi_0\rangle$ or $|\psi_1\rangle$.

Recall that a crucial assumption of the theorem is \emph{preparation independence}.  
If two systems are prepared independently, one in $|\psi_i\rangle$ and one in $|\psi_j\rangle$, then their joint ontic state $(\lambda_1,\lambda_2)$ is sampled from the product distribution:
\begin{equation} 
\mu_{\psi_i\psi_j}(\lambda_1,\lambda_2)
    = \mu_{\psi_i}(\lambda_1)\,\mu_{\psi_j}(\lambda_2).
\end{equation}
Hence, if the single-system distributions overlap, the region:
\begin{equation} \label{Sup}
\Lambda_\ast 
  \:= \mathrm{supp}(\mu_{\psi_0}) \cap \mathrm{supp}(\mu_{\psi_1}),    
\end{equation}
has nonzero probability, and therefore the product region
$\Lambda_\ast \times \Lambda_\ast$ is also reached with nonzero 
probability when the two systems are prepared independently.

In equation \eqref{Sup}, $\mu_{\psi_0}$ and $\mu_{\psi_1}$ are the epistemic (ontic) probability distributions associated with the quantum states $|\psi_0\rangle$ and $|\psi_1\rangle$, respectively.  
The notation $\mathrm{supp}(\mu_{\psi})$ denotes the \emph{support} of the distribution $\mu_{\psi}$, i.e.\ the set of ontic states $\lambda$ to which the preparation of $|\psi\rangle$ assigns nonzero probability.
Thus $\Lambda_\ast $ \eqref{Sup} is the set of ontic states that can arise from \emph{either} preparation $|\psi_0\rangle$ or $|\psi_1\rangle$ with nonzero probability.  
In a $\psi$-epistemic model, $\Lambda_\ast$ is nonempty, expressing the possibility that the two quantum states overlap in the underlying ontic space.

At any ontic state $(\lambda_1,\lambda_2)$ in this region, the model must predict the same measurement response probabilities \emph{regardless} of which of the four quantum product states was prepared.

This is where the PBR measurement enters.  
Suppose, \emph{as the $\psi$-epistemic model requires}, that the same ontic
state $(\lambda_1,\lambda_2)\in\Lambda_\ast\times\Lambda_\ast$ can be produced
with nonzero probability by \emph{each} of the four quantum preparations
$|\psi_i\psi_j\rangle$.  
Quantum mechanics predicts that each of the four product states
$|\psi_i\psi_j\rangle$ assigns probability zero to a \emph{different} measurement outcome.  Explicitly:

\begin{align*}
&|\psi_0\psi_0\rangle \;\Rightarrow\; p(0,0)=0, 
|\psi_0\psi_1\rangle \;\Rightarrow\; p(0,1)=0,\\
&|\psi_1\psi_0\rangle \;\Rightarrow\; p(1,0)=0,
|\psi_1\psi_1\rangle \;\Rightarrow\; p(1,1)=0.    
\end{align*}
Thus, each preparation forbids one outcome, and the four forbidden outcomes are all different.

Now assume that the same ontic state $(\lambda_1,\lambda_2)$ can arise from \emph{any} of these four preparations, as a $\psi$-epistemic model with overlap requires.  
In an ontological model, the quantum state $\psi$ does \emph{not} directly determine measurement probabilities. Instead, measurement outcome probabilities depend only on the ontic state $\lambda$. Thus, once the system is in ontic state $\lambda$, the measurement probabilities are fixed uniquely by $\lambda$, and cannot depend on which quantum state was prepared. 

Now, suppose a $\psi$-epistemic model claims that the same ontic state $(\lambda_1,\lambda_2)$ can arise from $|\psi_0\psi_0\rangle$ or $|\psi_0\psi_1\rangle$ or $|\psi_1\psi_0\rangle$ of $|\psi_1\psi_1\rangle$. Then the ontic state is the same physical reality in all four cases. The measurement device only “sees” the ontic state, not the quantum label attached to the preparation. So its output distribution must be the same:
\begin{equation*}
\xi((0,0)\mid \lambda_1,\lambda_2), \,\xi((0,1)\mid \lambda_1,\lambda_2), \, \xi((1,0)\mid \lambda_1,\lambda_2), \,
\xi((1,1)\mid \lambda_1,\lambda_2).    
\end{equation*}
These are fixed numbers, determined solely by $(\lambda_1,\lambda_2)$, not by $|\psi_i\psi_j\rangle$. Thus, we get the same $\lambda$ and so the probabilities for all four outcomes must be identical:
\begin{equation*}
\xi((0,0)\mid \lambda_1,\lambda_2)
=\xi((0,1)\mid \lambda_1,\lambda_2)
=\xi((1,0)\mid \lambda_1,\lambda_2)
=\xi((1,1)\mid \lambda_1,\lambda_2).    
\end{equation*}
If all four quantum preparations can produce the same ontic state $\lambda$, then at that ontic state the measurement device cannot “know” which preparation was used.
Since the measurement probabilities depend only on the ontic state $\lambda$, they must therefore be identical for all four preparations. 

However, the quantum predictions listed above require that, \emph{depending on which quantum state was prepared, one of these four probabilities must be exactly zero}.  If the same ontic state is compatible with all four preparations, then the same number would have to be zero in four different ways, forcing \emph{all} four probabilities to be zero simultaneously. The contradiction arises because orthogonality forces each preparation to assign probability zero to a different outcome.
However, \emph{this is impossible}, since the probabilities of the four outcomes must sum to 1:  

\begin{equation}
\sum_{i,j=0}^{1} \xi\!\big((i,j)\mid \lambda_1,\lambda_2\big) = 1.    
\end{equation}
Hence, the assumption that the same ontic state can arise from all four preparations leads to a direct contradiction with the quantum predictions.  \emph{The $\psi$-epistemic overlap must therefore be ruled out}.

The contradiction shows that the epistemic distributions \emph{cannot overlap}:
\begin{equation}
\mathrm{supp}(\mu_{\psi_0}) \cap 
\mathrm{supp}(\mu_{\psi_1}) = \varnothing .
\end{equation}
Thus, the quantum state must be a \emph{physical} property of the system, not merely a state of knowledge.  
This is precisely the conclusion of the PBR theorem: 
distinct quantum states correspond to disjoint sets of underlying physical states, and hence the quantum state is ontic \cite{PBR}.

Under assumptions 1)-3), the statistical-mechanical analogy leads to predictions that contradict quantum theory. The $\psi$-epistemic picture (in the HS sense) therefore fails. After PBR, we learn that every ontic state $\lambda$ is compatible with at most one quantum state.
Hence, if one had access to the true underlying physical state $\lambda$, one would know which $\psi$ was prepared, because no other quantum state could have produced that same $\lambda$.

It is important to emphasize what PBR do \emph{not} reject. The $\psi$-epistemic picture, as defined by HS and analyzed by PBR, does \emph{not} deny the existence of underlying reality. On the contrary, it assumes that each system has a real physical state $\lambda$, just as classical statistical mechanics assumes real microscopic states. PBR also \emph{accept} the existence of such an underlying reality. What they reject is the specific combination of ontological realism (there exists a real state $\lambda$), and an epistemic interpretation of $\psi$ (distinct quantum states overlap over the same region of $\Lambda$).

Given the existence of an underlying ontic state, PBR show that the epistemic reading of the wavefunction must fail. If reality exists at the ontic level, then distinct quantum states must correspond to disjoint sets of underlying physical possibilities.

\subsection{The Logical Arrow in Einstein’s 1935 Argument and Its Relation to PBR}
\label{Einstein-PBR}

PBR make a specific conceptual point when discussing Einstein’s 1935 correspondence \cite{PBR}: their theorem targets one logical direction of \emph{non-uniqueness} between quantum states and underlying reality, whereas Einstein’s argument concerns the \emph{opposite} direction. Keeping these arrows distinct is essential.

PBR note that two different forms of \emph{non-uniqueness} may arise between $\psi$ and reality \cite{PBR}:

\begin{description}
\item[A)] \emph{Two real states $\;\longrightarrow\;$ one $\psi$}.  
A single wavefunction corresponds to multiple underlying realities.  
This is the familiar notion of incompleteness: $\psi$ is too coarse-grained to describe all physical details.  
PBR frame their result as excluding $\psi$-epistemic overlap in the HS sense, a kind of non-uniqueness that can be schematically captured by A

\item[B)] \emph{One real state $\;\longrightarrow\;$ two $\psi$}.  
They argue that this is Einstein’s 1935 concern.  
Under locality, the same real physical state of system $B$ may be represented by several distinct conditional wavefunctions, depending solely on which measurement is freely performed on the spatially separated system $A$.
\end{description}

According to PBR, Einstein’s 19 June 1935 letter to Schrödinger analyzes scenario B \cite{PBR}.  
His reasoning proceeds from the separation principle: operations on system $A$ cannot alter the real state of $B$.  
Yet different measurements on $A$ yield different conditional states for $B$.  
Thus, one and the same real state of $B$ must correspond to multiple wavefunctions:
\begin{equation} \label{Ein}
\boxed{\lambda_B \;\Rightarrow\; \{\psi^1_B, \psi^2_B, \ldots\}}.
\end{equation}
The nonuniqueness lies in the \emph{representation}, not in the underlying ontology.  
If $\psi$ were a complete physical state, then distinct choices at $A$ would instantaneously alter the real state of $B$, violating separability.  
Einstein therefore concludes, via \emph{reductio}, that $\psi$ cannot be a complete description.  
This argument is ontological, not epistemic: it concerns physical coordination between descriptions and real states, not knowledge or ignorance.

By contrast, PBR argue that their theorem examines the reverse mapping \cite{PBR}: it analyzes epistemic models in which the wavefunction does \emph{not} represent the physical state.  In such models, a preparation associated with $\psi$ yields not a unique ontic state but a probability distribution $\mu_\psi(\lambda)$ over an underlying ontic state
space \cite{PBR}:
\begin{equation} \label{PBR}
\boxed{\psi_i \otimes \psi_j \;\Rightarrow\; \mu_{\psi_i}(\lambda_1)\,\mu_{\psi_j}(\lambda_2).}
\end{equation}
PBR ask whether two distinct quantum states $\psi$’s may correspond to \emph{overlapping} epistemic distributions.
In this picture, the wavefunction describes incomplete
information about which ontic state $\lambda$ is actual. 

\emph{This picture is completely different from Einstein’s}. For Einstein, the real state $\lambda_B$ of system $B$ is unique and well defined, while the wavefunction is a possibly nonunique \emph{representation} of that real state. His argument therefore concerns \emph{ontological underdetermination} (one $\lambda$ represented by many $\psi$), whereas the PBR analysis addresses \emph{epistemic overlap} (one $\psi$ compatible with many possible $\lambda$). The two perspectives are opposites: Einstein treats $\psi$ as an incomplete \emph{description} of a definite $\lambda$, while PBR test models in which $\psi$ is an incomplete \emph{state of knowledge} about $\lambda$.  \emph{The PBR theorem shows that allowing such epistemic overlap contradicts the predictions of quantum mechanics when combined with the preparation-independence assumption}.

Because Einstein emphasizes non-uniqueness, his view has occasionally been interpreted as a precursor to a modern $\psi$-epistemic model.  
But the Schrödinger–Popper correspondence shows otherwise.
Einstein does not give an epistemic interpretation of $\psi$ in the HS/PBR sense, as shown above.
This places his reasoning outside the HS and PBR framework.  
Einstein’s 1935 analysis lacks the structural components required by HS/PBR models:

\begin{enumerate}
    \item \emph{No ontic state space $\Lambda$.}  
    Einstein never posits a measurable space of hidden ontic states.  
    His “real state" is not a microstate variable drawn from an underlying probability distribution but a metaphysical state of affairs constrained by separability.
    
    \item \emph{No epistemic distributions $\mu_{\psi}(\lambda)$.}  
    Einstein does not represent $\psi$ as incomplete information about $\lambda$.  
    He never introduces the distributions required to define HS/PBR overlap.
    
    \item \emph{No probabilistic or Bayesian reading of $\psi$.}  
    His argument concerns ontological \emph{non-supervenience} of $\psi$ on reality, not degrees of belief or informational updating.
    \end{enumerate}

PBR explicitly acknowledge this contrast \cite{PBR}: they invoke Einstein not as a proponent of a $\psi$-epistemic model, but to illustrate the opposite kind of non-uniqueness.  
Their terminology is applied to Einstein only for limited comparative purposes.  
Einstein’s framework neither presupposes nor supports the structural assumptions required by the HS/PBR taxonomy.

\section{Ben{-}Menahem’s PBR Framework and the Einstein--Popper Correspondence} \label{3}

Yemima Ben{-}Menahem’s engagement with the PBR theorem, developed between 2015 and 2020 \cite{YBM-3, YBM-4, YBM, YBM-1, YBM-2}, provides the conceptual background for her interpretation of Einstein’s correspondence with Popper. The present section reassesses that interpretation in light of Einstein’s 1935 letters to Schrödinger and Popper, and in light of the logical contrast formalized in equations \eqref{Ein}--\eqref{PBR}. It also incorporates Ben{-}Menahem’s own crucial distinction (2017) between two different meanings of “$\psi$-epistemic,” a distinction that becomes methodologically decisive for how Einstein can, and cannot, be categorized.

\subsection{Two Epistemic Traditions in Ben{-}Menahem’s Framework} \label{Yem}

In her 2017 article, "The PBR theorem: Whose side is it on?" Ben{-}Menahem argues that the label "$\psi$-epistemic" actually covers two fundamentally different interpretive families. 
She distinguishes \cite{YBM}:

\begin{enumerate}
    \item \emph{Realist epistemic interpretations}.  
    These assume that a system possesses a determinate underlying physical state 
    $\lambda$, and take the quantum state $\psi$ to represent incomplete information
    about that underlying reality.

    \item \emph{Radical epistemic interpretations}.  
    These reject the existence of an ontic state space altogether: there are \emph{no}
    underlying physical states $\lambda$ for $\psi$ to describe.  
    Instead, $\psi$ is an informational or probabilistic tool---a "catalogue of expectations" in Schrödinger's phrase---for organizing the outcomes of possible
    measurements.
\end{enumerate}
Schrödinger wrote to Hans Berliner on 25 July 1935: "The $\psi$-function is designated and explained as a catalogue of expectations" (\emph{Erwartungskatalog}) \cite{Me}.

\medskip

Ben{-}Menahem stresses that the PBR theorem constrains \emph{only} the first family. 

Recall that the PBR framework \emph{presupposes}, as a starting assumption \cite{PBR}:

\begin{enumerate}
    \item \emph{An ontic space $\Lambda$ exists}, i.e,  PBR start by postulating that every quantum system has some underlying physical state $\lambda$, which is taken to be real, objective, and existing independently of observers. They start by postulating that every physical system has a real state $\lambda \in \Lambda$, the ontic state space.
    \item Once PBR assume that real states $\lambda$ exist, they impose the second assumption, \emph{the HS framework}. A preparation does not fix the real state. Still, it produces a distribution of possible real states: every laboratory preparation procedure $P_\psi$ (i.e., preparing a system in quantum state $\psi$) is represented by a probability distribution $\mu_\psi(\lambda)$ over $\Lambda$. The starting point in the PBR framework is the primitive mapping $P_\psi \rightarrow \mu_\psi(\lambda)$.
\end{enumerate}

The \emph{radical epistemic interpretation}, by definition, rejects both ideas and thus this ontic-level framework entirely. There is no hidden underlying physical state $\lambda$ in the first place. So \emph{there is no $\Lambda$}. Therefore, a preparation cannot be represented by \emph{any probability distribution over $\Lambda$}. That representation assumes precisely the kind of underlying reality Ben{-}Menahem rejects.
Thus, the radical epistemic interpretation cannot even be formulated within the PBR framework.
Therefore, Ben{-}Menahem concludes: "Once the distinction is in place, it becomes clear that the theorem targets the former variant of the epistemic interpretation, not the radical version.  
Rumors about the death of the epistemic interpretation were exaggerated" \cite{YBM}.

\subsection{Ben{-}Menahem’s Einstein Classification}
\label{misfires}

My claim in this subsection is that, by Ben{-}Menahem’s own realist/radical epistemic distinction, Einstein’s 1935 view does not straightforwardly fall within the HS/PBR class of $\psi$-epistemic models; any such classification requires importing formal structure into Einstein’s position that his texts do not supply.

\emph{The main difficulty in Ben{-}Menahem’s reading is not her taxonomy of contemporary epistemic views, but her historical placement of Einstein within that taxonomy}. 

In several works from 2015--2020 \cite{YBM-3, YBM, YBM-1, YBM-2} she links Einstein to an "ensemble/epistemic" tradition by reading his exchange with Popper as endorsing a statistical construal of Heisenberg uncertainty. She writes: "Evidently, Einstein did not see the uncertainty relations as applying to individual systems; in accordance with the ensemble interpretation, he took the values of conjugate observables to be well
determined, though not directly measurable" \cite{YBM}. 
On that basis, she treats Einstein as an early representative of the $\psi$-epistemic stance. As I argue below, \emph{this classification is placed under significant pressure by her own distinctions between realist and radical epistemic interpretations}:

\emph{1) Ensemble vs.\ $\psi$-epistemic are historically different questions.}
Ben{-}Menahem is right to stress that early foundational debates were not framed in HS/PBR terms.  
The ensemble interpretation is a claim about \emph{the referent of $\psi$}: it says $\psi$ describes statistical regularities in large ensembles, not the physical state of an individual system.  
HS and PBR, by contrast, take for granted that $\psi$ is a state-description of \emph{individual} systems and ask a different question: whether $\psi$ is ontic or epistemic \emph{with respect to an underlying ontic state space}.  
These debates belong to distinct conceptual landscapes.  
Reading Einstein as "$\psi$-epistemic" therefore retrofits a contemporary classifier onto a 1930s dispute whose axis was \emph{individual vs.\ ensemble}, not \emph{ontic vs.\ epistemic}.

Indeed, Ben–Menahem herself explicitly notes that neither the Copenhagen interpretation nor the early ensemble interpretation was "epistemic" in the modern HS/PBR sense; the epistemic/ontic divide is a later re-framing that should not be read back uncritically into the 1930s debates \cite{YBM-2}.

\emph{2) By Ben{-}Menahem’s own criterion, Einstein is not straightforwardly inside HS/PBR.}
In her 2017 distinction between \emph{realist epistemic} and \emph{radical epistemic} interpretations \cite{YBM}, Ben{-}Menahem insists that PBR constrains only the realist epistemic family: those that postulate an underlying ontic state space $\Lambda$ and treat $\psi$ as incomplete information about $\lambda\in\Lambda$ (see section \ref{Yem}).  
Radical epistemic views drop this postulate altogether and so fall outside HS/PBR from the outset; this is precisely why she exempts Itamar Pitowsky.%
\footnote{The PBR theorem is a no-go
result \emph{within} the HS class of realist--epistemic models.  Pitowsky’s interpretation does not belong to that class. Hence, the theorem cannot even be formulated against it, let alone refute it.
According to Ben{-}Menahem, Pitowsky rejects the HS/PBR starting point.  He denies the very existence of an ontic state space $\Lambda$, and thus also denies that probability distributions $\mu_\psi(\lambda)$ can represent preparations over such a space. With no $\Lambda$, there are no supports to overlap or fail to overlap, and no sense in which one could ask whether distinct quantum states correspond to the same underlying reality. He is therefore not
subject to the PBR result; his interpretation lies outside the theorem’s target class \cite{YBM, YBM-2}.}

However, once this distinction is in place, it becomes clear that Einstein’s 1935 reasoning likewise does not \emph{explicitly} instantiate an HS realist epistemic model.
His arguments (his letters to Schrödinger and Popper, see section \ref{1}) are framed in terms of separability, locality, and the coordination between real states and their quantum descriptions. He does \emph{not} postulate, nor even conceptualize, an underlying ontic state space $\Lambda$ equipped with epistemic probability distributions $\mu_\psi(\lambda)$, which is the structural core of the HS/PBR framework. 
Unlike Pitowsky, Einstein clearly insists on an underlying "real state of affairs." Still, he does not articulate this realism in the ontological-models vocabulary: there is no explicitly defined $\Lambda$, no family of preparation measures $\mu_\psi(\lambda)$, and no overlap structure.

Einstein, therefore, cannot simply be read as a \emph{realist epistemic} theorist in the HS technical sense; placing him in that category requires a substantive reconstruction that imports additional formal structure into his view. By the same structural criterion that excludes Pitowsky from the PBR target class, Einstein’s 1930s analysis is at best only indirectly representable within the HS/PBR framework, rather than a straightforward instance of it.
This is not to deny that one may retrospectively represent Einstein’s position within an ontological-models framework; many contemporary authors do precisely that, treating $\Lambda$, $\mu_{\psi}(\lambda)$, and overlap as useful reconstructions rather than historical commitments. My point is not that such reconstruction is illegitimate, but that Einstein’s own texts do not supply it—and that Ben-Menahem herself insists on precise structural criteria when excluding Pitowsky from the HS/PBR class. If those criteria are applied symmetrically, Einstein’s 1935 view likewise lies outside the HS/PBR category unless significant formal machinery is imported into it.

\emph{3) A selective asymmetry in application.}
The resulting asymmetry is methodological rather than purely historical: Pitowsky is placed \emph{outside} HS/PBR because he posits no $\Lambda$, yet Einstein is drawn \emph{inside} the epistemic side of the HS/PBR divide even though his texts do not supply any $\Lambda$-based epistemic structure.  
This selective extension of the modern classifier suggests that contemporary categories are being retrofitted to Einstein’s position in a way that risks outstripping the principled taxonomic constraints Ben{-}Menahem herself introduces.

\subsection{An Ensemble Theorist but not a \texorpdfstring{$\psi$}{psi}-Epistemic Theorist}

A further difficulty in assessing Ben{-}Menahem’s placement of Einstein within an epistemic lineage concerns the textual basis of her interpretation. Ben{-}Menahem follows Popper’s English rendering of Einstein’s 1935 letter, but the German manuscript reveals a more cautiously framed argument. 
Popper’s translation reflects his own statistical program: it aligns Einstein with Popper’s own statistical reading of the uncertainty principle and with the interpretation Popper was defending at the time.
Popper’s version smooths out Einstein’s more cautious, qualified remarks and introduces explicitly probabilistic language that is absent in the original.

The relevant passage reads:%
\footnote{The literal English translation from the German is my own. Einstein to Popper, 1935; German letter reproduced in \cite{Popper}:
\begin{quote}  
“Man kann sich fragen, ob der statistische Charakter unserer experimentellen Befunde im Heutigen Quantenforschens erst durch die fremden Eingriffe während Messungen verursacht wird, während die Systeme an sich, beschrieben durch eine $\psi$-Funktion, deterministisch sich verhalten. (Heisenberg liebäugelt mit einer solchen Auffassung, ohne sie konsequent zu vertreten.)
Man kann auch so fragen: Ist das die $\psi$-Funktion, deren zeitliche Abänderung durch die Schrödinger Gleichung zeitlich deterministisch verläuft, wirklich als vollständige Beschreibung des physikalischen Zustandes eines Systems anzusehen, wobei lediglich die fremden Eingriffe während Beobachtungen dafür verantwortlich sind, dass die Aussagen nur statistischen Charakter haben?
Die Antwort ist, dass die $\psi$-Funktion nicht als vollständige Beschreibung des Zustandes aufgefasst werden darf.”
\end{quote}}

\begin{quote}
One may ask whether the statistical character of our experimental findings in present-day quantum research is produced solely by external intervention during measurement, while the systems themselves, described by a $\psi$-function, behave deterministically. (Heisenberg flirts with such a conception, without adopting it consistently.) One may also pose the question as follows: Is the $\psi$-function—whose temporal change proceeds deterministically according to the Schrödinger equation—to be regarded as a complete description of the physical state of a system, with the merely external interventions during observation being solely responsible for the fact that the predictions have only a statistical character?
The answer is that the $\psi$-function must not be regarded as a complete description of the state.
\end{quote}

Popper’s translation remains generally close to the meaning, but it introduces several subtle shifts. His rendering softens Einstein’s reference to “present-day quantum research,” adds a normative tone (“should we regard...”), and expands “fremde Eingriffe” into “(insufficiently known) interference,” thereby \emph{importing epistemic vocabulary not present in the German text}. The cumulative effect is to make Einstein’s argument appear more closely aligned with Popper’s statistical–epistemic program than the original wording requires. Einstein’s own phrasing is sparser, more structural, and framed as a conceptual question about completeness rather than about epistemic interference.

This distinction matters because interpretations based solely on Popper’s English version may overemphasize the epistemic–probabilistic dimension of Einstein’s remarks.

Later, in his 1967 paper, "Quantum Mechanics without 'The Observer',” Popper presents Einstein as advancing a realist, ensemble-based interpretation whose conceptual structure is best characterized as statistical-epistemic rather than ontological \cite{Pop}: 
First, Popper ties Einstein directly to an ensemble-based interpretation of quantum mechanics: 
"There is the German physicist, Fritz Bopp, who ... develops ... a theory with which Einstein would hardly have had any quarrel since, on lines not dissimilar to Einstein's ... Bopp interprets the quantum theoretical formalism as an extension of classical statistical mechanics; that is, as a theory of ensembles."    
Second, Popper reads Einstein as giving a statistical, rather than ontological, interpretation of the wavefunction: "Yet as Einstein had offered his own (statistical) interpretation of quantum theory, he clearly accepted its consistency." Third, Popper reads Einstein as a realist who holds that physical properties exist independently of the wavefunction: "Thus the so-called 'paradox' of Einstein, Podolsky, and Rosen ...is not a paradox but a valid argument, for it established just this: that we must ascribe to particles a precise position and momentum, which was denied by Bohr and his school..." \cite{Pop}.

In the German text of Einstein's letter to Popper, Einstein’s focus falls on the logical coordination between (1) the deterministic Schrödinger evolution and (2) the ontological completeness of the state description. His conclusion—that $\psi$ “must not be regarded as a complete description of the state”—is \emph{an ontological} claim about insufficient descriptive resources, \emph{not} a probabilistic thesis about information, ignorance, or epistemic accessibility.

Read directly from the German, Einstein’s position corresponds to what would later be called the ensemble interpretation. Several points follow explicitly:

\begin{enumerate}
    \item The $\psi$-function does not describe the physical state of an individual system.
    \item The statistical character of predictions reflects the use of $\psi$ over ensembles 
    
    \noindent (“Aussagen nur statistischen Charakter”).
    \item A complete description requires additional elements not contained in $\psi$ 
    
    \noindent (“nicht als vollständige Beschreibung…aufzufassen”).
\end{enumerate}

These commitments place Einstein within the historical ensemble tradition and help explain why later authors sometimes read his view as exhibiting certain $\psi$-epistemic features \cite{Boge, Le}.
However, in the HS/PBR technical sense, $\psi$-epistemic models require specific structural machinery, and Einstein’s 1930s writings supply none of these components.
\emph{Einstein’s reasoning lacks the structural characteristics that define the HS/PBR class of epistemic models}: there is no measurable ontic space $\Lambda$, no epistemic distributions $\mu_{\psi}(\lambda)$, and no overlap structure. His argument concerns locality, separability, and descriptive completeness—not epistemic probability over hidden variables.
\emph{Einstein’s conclusions do not resemble modern $\psi$--epistemic models in their HS/PBR technical structure; his remark that the $\psi$-function is “\emph{nicht als vollständige Beschreibung}” marks an ontological limitation of the formalism, not an epistemic one.}

\subsection{Broad Epistemic Resonances, No HS/PBR Structure}

In the HS/PBR technical sense, Einstein’s conclusions do not resemble modern $\psi$-epistemic ontological models. However, the gap between his ensemble view and contemporary $\psi$-epistemic approaches becomes significant at the points where Ben{-}Menahem juxtaposes Einstein’s ensemble view with the HS/PBR epistemic category, drawing on their shared structural features while maintaining that Einstein’s interpretation \emph{is not itself} $\psi$-epistemic in the HS/PBR sense.

Once the German text is taken into account, Einstein’s stance emerges as ensemble-based and ontological rather than epistemic in the HS/PBR sense. Ben{-}Menahem never explicitly states that Einstein is a $\psi$-epistemic theorist; her claim is more nuanced. She presents Einstein’s ensemble interpretation as historically continuous with later epistemic approaches at the level of broad interpretive attitude, while also marking a principled divergence from the radical epistemic interpretations that dispense with underlying definite states altogether.

The appearance of continuity rests on several structural and conceptual affinities that she draws between Einstein’s view and the epistemic models targeted by PBR \cite{YBM}:

Both the ensemble interpretation and the non-radical epistemic interpretations agree in denying that the quantum state is a unique representation of the physical state of an individual system. Ben{-}Menahem emphasizes that, in the early debates, the ensemble view “understand[s] QM as a full-blown probabilistic theory, that is, a statistical description of an \emph{ensemble} of similar systems that makes no definite claims about individual members of the ensemble.” In this picture, the wavefunction does not describe the real state of a single particle but only encodes the statistical behavior of an ensemble. She later notes that the epistemic interpretation, as defined by HS, “has much in common with the earlier ensemble interpretation; neither of them construes quantum states as uniquely representing physical states of individual systems, and both can dismiss the worry about the collapse of the wavefunction.” The first strand of continuity, then, lies in the shared rejection of a direct ontic reading of the wavefunction at the level of the individual system \cite{YBM}. 
According to Ben{-}Menahem, HS’ characterization of $\psi$-epistemic models—where different quantum states correspond to overlapping probability distributions over the same underlying physical state space—makes explicit what, in the historical ensemble picture, is often left at the heuristic level. Ben{-}Menahem underlines this when she observes that the “raison d’être of the ensemble interpretation” is precisely the assumption of an underlying level of well-determined physical states; non-radical epistemic views reinterpret this assumption in information-theoretic terms \cite{YBM}.

\subsection{By Ben{-}Menahem's Own Distinction Einstein Lies Outside HS/PBR}

My point is not that Ben-Menahem explicitly mislabels Einstein in a straightforward taxonomic sense. Instead, the difficulty arises from her own distinction between \emph{realist epistemic} and \emph{radical epistemic} interpretations—introduced expressly to avoid retroactively projecting modern classifications onto historical figures.
The issue is that Ben-Menahem's distinction between \emph{realist epistemic} and \emph{radical epistemic} interpretations reveals that Einstein’s 1935 argument does not supply the formal assumptions required to place him \emph{within} the HS/PBR framework. 

Ben{-}Menahem explicitly uses this distinction to shield Pitowsky from the scope of the PBR theorem: Pitowsky posits no underlying ontic state space $\Lambda$, no epistemic probability distributions $\mu_\psi(\lambda)$, and no overlap structure \cite{YBM-2}. 
Yet Einstein likewise provides none of these structural elements. His argument contains no $\Lambda$-space, no epistemic distributions, no notion of overlapping supports, and no interpretation of $\psi$ as information about hidden microstates. It concerns locality, separability, and the completeness of physical description, not epistemic probability. 

Einstein has neither the HS/PBR structure nor the radical epistemic structure. So by Ben{-}Menahem's own distinction, Einstein lies outside HS/PBR, just as Pitowsky does.
This does not mean that Einstein’s view cannot be reconstructed within an ontological-models framework; it means only that, on Ben-Menahem’s own criteria, such a reconstruction is an additional interpretive step rather than something supplied by Einstein’s texts themselves.
Consequently, the very criterion Ben{-}Menahem invokes to exempt Pitowsky from HS/PBR models also exempts Einstein. The classification is thus placed under significant pressure. If Pitowsky lies outside HS/PBR for lacking the required formal architecture, then Einstein must fall outside it for the same reason. Put differently, \emph{if Pitowsky is exempt because he offers no $\Lambda$, no $\mu_\psi(\lambda)$, and no overlap structure, then Einstein must be exempt for the very same reason}.

By Ben{-}Menahem's own standards, Einstein's 1935 view does not supply the structural elements required for classification within the HS/PBR space of models. The HS framework requires 1) a well-defined ontic state space $\Lambda$, 2) preparation distributions $\mu_{\psi}(\lambda)$, and 3) the possibility of overlapping supports. Einstein's ensemble interpretation provides none of these. It asserts only that $\psi$ fails to describe the individual physical state and that statistical predictions arise from its ensemble use. This is an ontological incompleteness claim, not an epistemic--probabilistic model over a hidden state space. Thus, \emph{mutatis mutandis}, the very criteria Ben{-}Menahem invokes to exempt Pitowsky from HS/PBR also exempt Einstein. Her historical classification of Einstein as belonging to the $\psi$-epistemic lineage therefore depends on importing structural assumptions into Einstein's position that his own texts do not contain.

\subsection{Objections Raised by a Defender}
\label{defense}

A defender of Ben{-}Menahem might raise several objections.  
None of them ultimately removes the structural tension identified above; instead, they help clarify why caution is needed when mapping HS/PBR categories onto Einstein’s 1930s position.

\emph{Objection 1: "Einstein assumed underlying realities, so HS/PBR can apply."}
Indeed, Einstein posits \emph{real states}.  
But HS/PBR realism is \emph{not} simply the claim that reality exists.  
The HS/PBR framework requires a specific formal ontology: a measurable ontic state space $\Lambda$, and epistemic probability distributions $\mu_\psi(\lambda)$ over $\Lambda$ for each preparation.  
Einstein’s realism is articulated in terms of \emph{separability and locality}, not in terms of a statistical ontology over hidden-variable state spaces.  
He never posits an HS-style “probability-over-$\Lambda$" framework, nor anything resembling the ontological-models structure on which HS/PBR rests.
Thus, while Einstein’s commitment to realism is clear, the formal machinery to which PBR applies is not supplied by his view.

\emph{Objection 2: "Ben{-}Menahem is using ‘epistemic’ in a loose historical sense, not an HS/PBR sense."}
That is precisely the interpretive difficulty.  
If “epistemic" is used historically, one should not import HS/PBR consequences into the Einstein case.  
If it is used with HS/PBR precision, then Ben{-}Menahem’s own 2017 distinction between \emph{realist} and \emph{radical} epistemic views immediately shows that Einstein does not exhibit the formal structure defining the realist epistemic class.  
Equivocation on the meaning of “epistemic" creates the impression of continuity between Einstein and HS/PBR models, even though the relevant technical assumptions are absent.

\emph{Objection 3: “Einstein is closer to ensemble epistemicism than to $\psi$-ontology, so classifying him as epistemic is harmless}.”
At most, this yields a family resemblance, not membership in a taxonomic category.
One may grant that Einstein’s ensemble interpretation shares some surface-level affinities with later epistemic views, but this does not identify his position with HS/PBR epistemic models.
Such models presuppose an ontic space with epistemic probability distributions — a structure absent from Einstein’s “nicht als vollständige Beschreibung” argument, which is explicitly ontological.
Similarity in motivation or terminology does not entail equivalence in ontology.

\medskip

The foregoing analysis also clarifies the status of the PBR theorem.
Many responses to PBR—including Ben-Menahem’s realist/radical epistemic distinction—aim to limit its scope by showing that certain epistemic readings of the quantum state fall outside its target class.
Yet once the structural assumptions of the HS/PBR framework are applied symmetrically, this strategy proves less effective than often claimed: models that evade PBR typically do so only by lacking the hidden-variable architecture the theorem presupposes, rather than by providing a viable epistemic alternative.
In this sense, the present analysis indirectly reinforces the PBR theorem’s force as a genuine no-go result for epistemic hidden-variable models.

\end{document}